\documentclass[preprint]{aastex}

\usepackage{graphicx}
\usepackage{dcolumn}
\usepackage{bm}
\usepackage{epstopdf}

\newcommand{\myskip}[1]{}
\newcommand{\iso}{{\rm iso}}

\renewcommand{\d}{{\rm d}}

\newcommand{\gb}{{\bar g}}

\newcommand{\BEQ}{\begin{eqnarray}}
\newcommand{\EEQ}{\end{eqnarray}}
\newcommand{\BEA}{\begin{eqnarray}}
\newcommand{\EEA}{\end{eqnarray}}
\newcommand{\nn}{\nonumber}

\newcommand{\eV}{{\rm eV}}

\shorttitle{Neutrino mass from lensing in A1689}

\shortauthors{Nieuwenhuizen and Morandi}

\begin{document}

\title{Prediction for the neutrino mass in the KATRIN experiment from lensing by the galaxy cluster A1689}
\author{Theo M. Nieuwenhuizen}
\affil{ Center for Cosmology and Particle Physics, New York University, New York, NY 10003\\
On leave of:  Institute for Theoretical Physics, University of Amsterdam,
Science Park 904, \\ P.O. Box 94485, 
1090 GL  Amsterdam, The Netherlands}

\author{Andrea Morandi}
\affil{Raymond and Beverly Sackler School of Physics and Astronomy, Tel Aviv University, Tel Aviv 69978, Israel}

\email{t.m.nieuwenhuizen@uva.nl;  andrea@wise.tau.ac.il}

\begin{abstract}
The KATRIN experiment in Karlsruhe Germany will monitor the decay of tritium, which produces an electron-antineutrino. 
While the present upper bound for its mass is 2 eV/$c^2$, KATRIN will search down to 0.2 eV$/c^2$. If the dark matter of the galaxy 
cluster Abell 1689 is modeled as degenerate isothermal fermions, the strong and weak lensing data may be explained by degenerate 
neutrinos with mass of 1.5 eV$/c^2$. Strong lensing data beyond 275 kpc put tension on the standard cold dark matter interpretation.
In the most natural scenario, the electron antineutrino will have a mass of 1.5 eV/$c^2$, a value that will be tested in KATRIN.
\end{abstract}


\maketitle

\section{Introduction}

The neutrino sector of the Standard Model of elementary particles is one of today's most intense research fields 
(Kusenko, 2010; Altarelli and Feruglio 2010; Avignone et al. 2008). While neutrino oscillations give information of differences 
in mass-squared values, no absolute neutrino mass is known. The masses may be of order 0.05 eV/$c^2$ as set by the oscillations, 
or much larger. While a lot is known about standard (ÒactiveÓ) neutrinos (left-handed neutrinos, right-handed antineutrinos), 
hardly anything is known about their ÒsterileÓ partners (right-handed neutrinos, left-handed antineutrinos). 
It is even not established that the masses of active and sterile components are close to each other; in the popular seesaw mechanism, 
they differ by orders of magnitude (Lesgourges and Pastor 2006). 
Still, next to WIMPs (Bertone 2010), sterile neutrinos may constitute 
the dark matter particles (Feng 2010). It was recently realized that lensing data allow that active plus sterile neutrinos of 1.5 eV/$c^2$ 
provide the dark matter (Nieuwenhuizen et al. 2009). Neutrinos are too light to trigger 
large scale structure formation (galaxy formation), and are therefore often ruled out as dark matter candidates. 
However, this may end up as a misconception, since structure formation
 can actually already be explained by the gravitational hydrodynamics of baryons alone (Nieuwenhuizen et al. 2009).

The most effective way to determine the mass of an antineutrino is tritium decay, when one of the two neutrons in its nucleus 
decays into a proton, an electron and an electron-antineutrino. The electron energy can be detected. Analyzing the number of events
 (that is, the Kurie function) near the maximal energy provides information about the antineutrino mass. Depending on statistical 
 assumptions in the analysis of the non-detection in the Mainz and Troitsk experiments, the present upper bound is 2.2 eV/$c^2$ 
 (Lobashev et al. 1999), 2.3 eV/$c^2$ (Kraus et al.  2005) or 2.0 eV/$c^2$ (Amsler et al. 2008). 
 
 The KArlsruhe TRItium Neutrino experiments (KATRIN) involve an international consortium of 16 laboratories in Europe 
 and the USA, and are intended to search the mass down to 0.2 eV/$c^2$ (Otten and Weinheimer 2009). Since the square mass 
 is tested, this means to go down to 0.04 eV$^2/c^4$, an improvement by two orders of magnitude with respect to Mainz-Troitsk.
 
Galaxy clusters are known to produce strong lensing effects. A galaxy lying behind it can be seen as an arc, 
 ideally as a complete Einstein ring. This has been observed, see Fig. 1 which presents a double Einstein ring.
 The very similar phenomenon may be seen already in lensing by a glass ball, see figure 2.
  This so-called strong lensing is effective near the center of the cluster, say up to a radius of 300 kpc.
   Further out, there is still the so-called weak lensing. Indeed, since every galaxy will be slightly distorted, 
  a statistical effect occurs for a ``localÓ group of galaxies (``localÓ on the 2D sky, as observed by us), which then also provides 
  information on the mass distribution of the lensing cluster. Models for structure of the cluster can be tested against these strong 
  and weak lensing data. The clusters also have an amount of 10-15\% of matter in X-ray gas, typically much more than they 
  have galaxy mass.
  
The galaxy cluster Abell 1689 (A1689) is one of the most studied clusters. It is rather heavy, a few times $10^{15}$ solar masses, 
  and has a large lensing arc at 150 $h_{70}^{-1}$ kpc, which is observed at 45'' from the center, the largest Einstein radius observed to date. 
  (The Hubble constant is $H_0 = h_{70}70$ km/s Mpc, with $h_{70}$ close to 1.0). 
  There are good data for strong lensing (SL) and weak lensing (WL) and the X-ray gas, so it is an ideal testing ground for models. 
  The cluster has a small intruder clump in the North East sector, which can be masked out if needed.
 Assuming spherical symmetry, we shall consider two models for the matter distribution in the cluster.

\begin{figure}[h!]
\centering
\includegraphics[width=12cm]{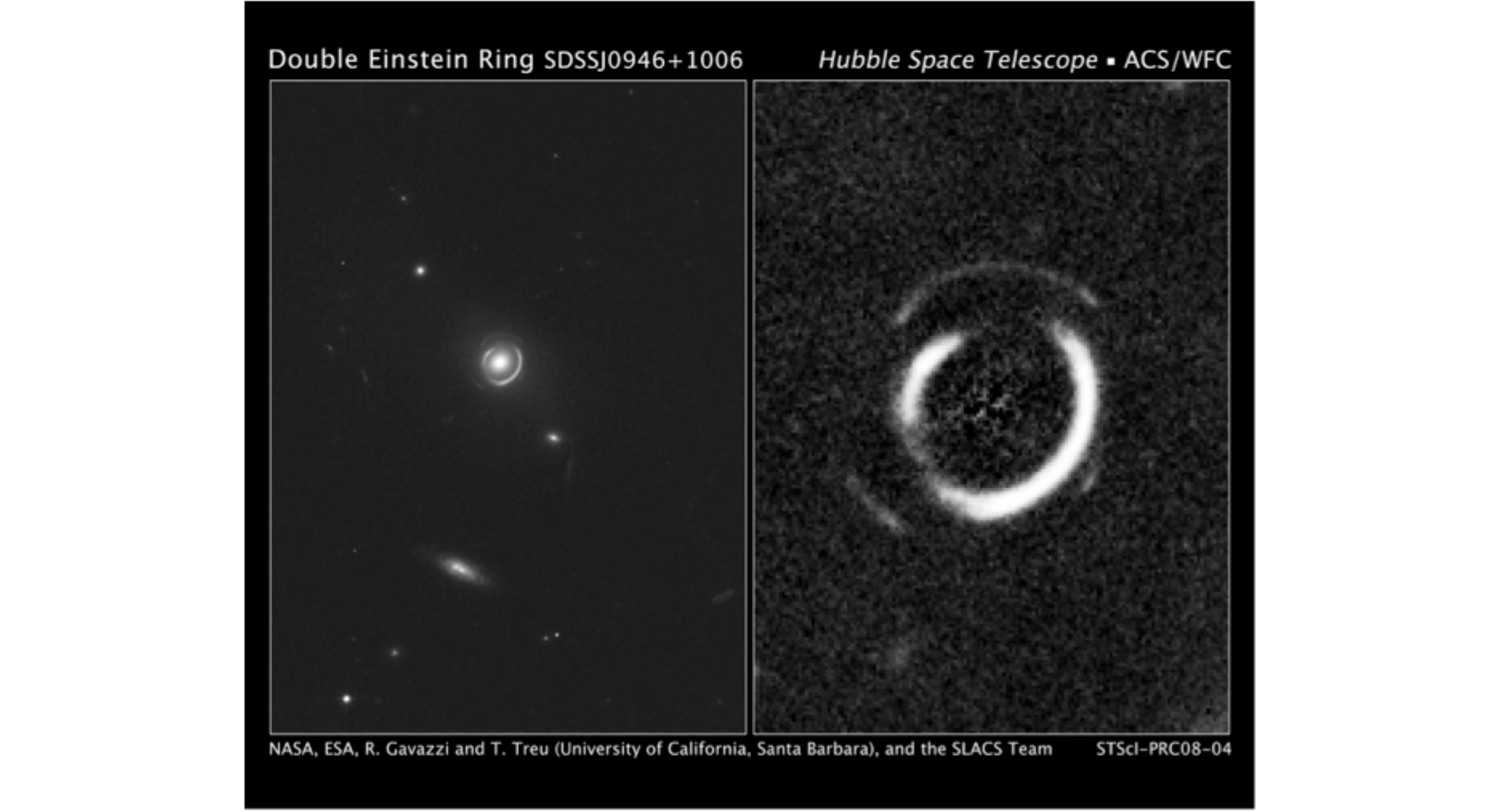}
\vspace{-0cm}
\caption{Strong lensing by a foreground galaxy photographed by the Hubble Space Telescope HST. 
Two background  galaxies are mapped as two almost complete Einstein rings. 
In the enlargement on the right, the foreground galaxy is masked out.}
\end{figure}


\begin{figure}[h!]
\centering
\includegraphics[width=5cm]{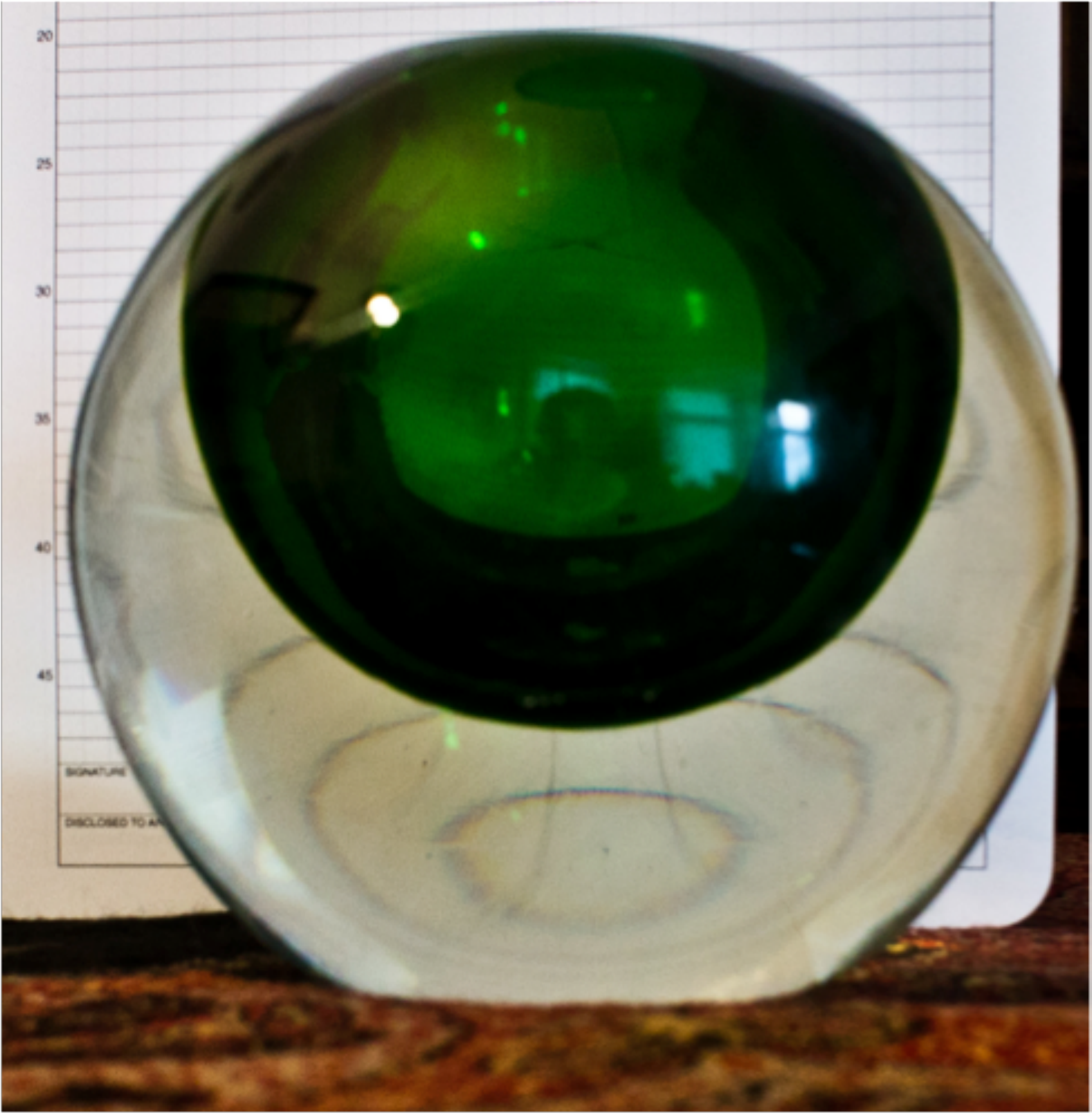},
\vspace{-0cm}
\caption{Strong lensing by a glass ball at home. The central square of the background paper is distorted. 
Parallel lines are mapped as semi-arcs,  pairs of which make up the ellipses. }
\end{figure}

 \section{NFW profile for cold dark matter} 
  
Cold dark matter (CDM) is mostly thought to be related to particles heavier than the proton, 
like the lightest supersymmetric (SUSY) particle called neutralino. However, the assumption of CDM is subject to criticism. 
The so-called R-symmetry assumption implies a very long lifetime. The electron and proton have spin, 
are both charged and light, so they ``have nowhere to go''; but if the R-symmetry is broken, the neutralino would quickly 
decay to electrons, protons, neutrinos and photons. In practice, the purported CDM particle has never been found, 
despite decades of searching. An incomplete list of acronyms of the over 60 search programs is, not mentioning updates:
ADMX, AMANDA, ANAIS, ANTARES, ArDM, ATIC, BPRS, CANGAROO, CAST, CDMS, CLEAN, CMSSM, CoGeNT, COUPP, 
CRESST, CUORE, CYGNUS, DAMA/NaI, DAMA/LIBRA, DAMIC, DEAP, DMTPC, DRIFT, EDELWEISS, EGRET, 
ELEGANTS, FERMI-LAT, GENIUS, GERDA, GEDEON, HDMS, ICECUBE, IGEX, KIMS, LEP, LHC, LIBRA, LUX, MAGIC, 
MAJORANA, MIMAC, NAIAD, NEWAGE, ORPHEUS, PAMELA, PICASSO, ROSEBUD, SIGN, SIMPLE, SuperCDMS, 
SUPER-K, TEXONO, UKDM, VERITAS, WArP, Whipple, XENON, XMASS, ZEPLIN. 
Upcoming are: DARWIN, DEAP-3600, EURECA, GEODM, MiniCLEAN, SuperCDMS-SNOLab.
(Notice that J, Q and Y are not covered.)
Also in recent years the searching remains without detection, in particular the XENON 100 search rules out 
previous detection claims at DAMA, CDMS and Cogent (Aprile et al. 2010),
a finding confirmed by the CDMS collaboration (Akerbib et al. 2010).
While it is already acknowledged that time for supersymmetry is running out  (Bertone 2010),
the CMS collaboration at the Large Hadron Collider LHC reports, very recently, that within error limits it 
fails to detect more SUSY-like events than expected from the non-SUSY background (CMS 2011). 
So, though the SUSY assumption and the possibly resulting neutralino CDM are certainly not ruled out,
these long trusted options are definitely on the defensive at the moment.

From the practical side in cosmology, the CDM assumption works often so well that it is presently viewed as the  
standard approach. This applies in particular for the cosmic microwave background. As to cosmic structures,
numerical simulations have shown that the CDM density can be described fairly well by an NFW profile 
(Navarro, Frenk and White, 1997), which has the benefit of having a simple form. It is parametrized by a 
characteristic mass density $\rho_0$ and a length scale  $r_s$,

\BEQ
\rho_{\rm NFW}(r)=\frac{\rho_0\,r_s^3}{r(r+r_s)^2}.
\EEQ
The average density, that is, the integrated mass up to $R$ divided by the volume, then equals 

\BEQ \bar\rho_{\rm NFW}(R)=\frac{3}{4\pi R^3}\,4\pi \rho_0r_s^3\,\left(\ln\frac{R+ r_s}{r_s}- \frac{R}{R + r_s}\right).
\EEQ

The cosmic critical density is $\rho_c=3H_0^2/8\pi G =9.25\cdot 10^{-27} $kg/m$^3$ according to
the WMAP7 value for the Hubble constant,  $H_0=70.2\pm1.4$ km/s/Mpc (Komatsu et al. 2010).
At the redshift $z=0.183$ of A1689 the Hubble constant reads $H_z\equiv E_z\,H_0$ with

\BEQ
E_z \!=\left[\Omega_M (1+z)^3 +  (1-\Omega_M-\Omega_{\Lambda})(1+z)^2 + \Omega_{\Lambda}\right]^{1/2}=1.078,
\EEQ
(we adopt the flat cosmology $\Omega_\Lambda=0.75$, $\Omega_M=0.25$), so the 
critical density becomes $\rho_c(z) =\rho_c E_z^2$.
The radius $R_N$ where $\bar\rho_{\rm NFW}(R_N)$ equals $N\rho_c(z)$ is said to be the radius where the over-density equals $N$. 
It defines the concentration parameter $c_N=R_N/r_s$.
Customary values are $N=200$ corresponding to $c_{200}$ and ``virial'' overdensity $N=117$ at the redshift of A1689,
with its $c_{\rm vir}\equiv c_{117}$.
The correspondence involved here is
\BEQ
\frac{\rho_0}{N\rho_c(z)}=\frac{\frac{1}{3}c_N^3}{\log(1+c_N)-c_N/(1+c_N)},
\EEQ
with the monotonically increasing right hand side yielding a unique $c_N$.

\section{Isothermal neutrino mass models}

  It is long believed that light neutrinos cannot make up the dark matter, because they cannot assist in structure formation
  of galaxies and clusters of galaxies.   Indeed, the Jeans mechanism in hydrodynamics can only explain globular star clusters of, say, 
  600,000 solar masses.  But gravitational hydrodynamics offers new mechanisms for cosmological structure formation, with an important role 
  for viscosity. We shall not dwell into it here, as it is discussed at length in other papers in this volume by Gibson and Schild (2011) and Gibson (2011).  
  This approach also answers the paradox why the $^3$He abundance in the Galaxy 
  is often observed at primordial level, independent of the the distance to the center and thus independent of the local
   amount of metallicity (Nieuwenhuizen, 2011).
  For the present work it suffices to notice that neutrinos need not be ruled out by arguments of structure formation
  and hence qualify ``again'' for dark matter candidates.
  This holds the more since the early runs of the Large Hadron Collider at the CERN did not detect supersymmetry  (CMS 2011).
  
One may thus consider the cluster dark matter as due to degenerate fermions. The violent relaxation mechanism has explained 
  the success of isothermal models in gravitation (Lynden Bell 1967), so it is natural to consider isothermal dark matter fermions. 
 

 The gravitational potential caused by mass elements at ${\bf r}'\neq{\bf r}$ is 
 $\varphi({\bf r})=-G\int\d^3r'\rho({\bf r}')/|{\bf r}-{\bf r}'|$ + $G\int\d^3r'\rho({\bf r}')/r'$, where we
 choose the zero level such that $\varphi({\bf 0})=0$. It satisfies the Poisson equation 
 $\nabla^2\varphi({\bf r})=4\pi G\rho({\bf  r})$, which in case of spherical symmetry reads

\BEQ
\varphi''(r)+\frac{2}{r}\varphi'(r)=4\pi G\rho(r), \quad
\EEQ
In the galaxy cluster there will be three relevant contributions to the matter density,
$ \rho=\rho_\nu+\rho_G+\rho_g$, with partial mass densities of dark matter (``$\nu$''),
 Galaxies (G) and X-ray gas (g), respectively.  We model the dark matter as isothermal fermions 
with mass $m$, degeneracy $\gb$ and  chemical potential $\mu=\alpha k_BT$. The mass density
then involves a momentum integral over a Fermi-Dirac distribution, 

\BEQ 
\rho_\nu(r)=\gb m\,\int\frac{\d^3p}{(2\pi\hbar)^3}\,\frac{1}{\exp\{[p^2/2m+m\varphi(r)-\mu]/k_BT\}+1}.
\EEQ
The thermal length is $\lambda_T=\hbar\sqrt{2\pi/mk_BT}$.

Since the mass in galaxies is a few percent of the total, mainly coming from the dominant central one, we assume for the 
mass density from galaxies simply the isothermal distribution  $\rho_G(r)=\gb m\lambda_T^{-3}\exp[\alpha_G-\beta_Gm\varphi(r)]$ 
(Nieuwenhuizen 2009).  In the galaxy cluster Abell 1689 the mass density of the gas $\rho_g$ follows from the electron density
 $n_e$ as $\rho_g=1.11\,m_N n_e$ with $m_N$ the nucleon mass. 
The data for $n_e$ densely obtained by us from X-ray observations in the interval $10$ kpc $<r<$ 1 Mpc,
are well approximated by the exponent of a parabola in $\log r$ (Nieuwenhuizen and Morandi, 2011)

\BEQ \label{electrondensity} \hspace{-0cm}
n_e(r)= \exp\left({
-3.63 + 0.915 \log \frac{r}{{\rm kpc}} - 0.218 \log^2\frac{r}{{\rm kpc}} }\right){h_{70}^{1/2}}{{\rm cm}^{-3}}.
\EEQ
Beyond 1 Mpc one may either continue this shape, or replace it for $r>r_{\rm iso}=807$ kpc by the isothermal tail 
(Nieuwenhuizen and Morandi, 2011)
\BEQ 
\rho_g^{\rm iso}=
\frac{\sigma_\iso ^2}{2\pi G r^2}, \qquad 
\qquad \sigma_\iso =578.5\,\frac{{\rm km}}{{\rm s}}.\EEQ

{} From strong lensing data one determines the $2D$ mass density
(integrated mass along the line of sight)
$\Sigma(r)=\int_{-\infty}^\infty\d z\rho\left(\sqrt{r^2+z{}^2}\,\right)$.
The Poisson equation allows to express this as

\BEQ \label{SigmaFromPhip}
\Sigma(r)=
\frac{1}{2 \pi G} \int_0^\infty\d s\frac{ \cosh 2s}{\sinh^2s}
\left[\varphi'(r\cosh s)-\frac{\varphi'(r)}{\cosh^2s}\right],
\EEQ
which is numerically  more stable since $\varphi'$ varies less than $\rho$.
In weak lensing one determines the shear as

\BEQ\label{gt=}
g_t(r)=\frac{\overline{\Sigma}(r)-\Sigma(r)}{\Sigma_c-\Sigma(r)},
\EEQ
where the ``critical projected mass density'' $\Sigma_c$ is a constant.
The mass inside a cylinder of radius $r$ around the cluster center, $M_{2D}(r)=2\pi\int_0^r\d r'\,r'\Sigma(r')$, defines 
the average of $\Sigma$ over a disk of radius $r$, 
$\overline{\Sigma}(r)=M_{2D}(r)/\pi r^2$. 
The Poisson equation allows to express this as (Nieuwenhuizen 2009) 

\BEQ \overline{\Sigma}(r)=
\frac{1}{\pi G} \int_0^\infty\d s\, \varphi'(r\cosh s).
\EEQ



 \section{Applications of the NFW and isothermal neutrino models}
 
  Cowsik and McClelland (1973) are the first to model the dark matter of the Coma cluster as an isothermal sphere of neutrinos. 
  As for neutron stars and white dwarfs, this determines the neutrino mass, and yields about 2 eV/$c^2$. Treumann et al. (2000) considered 
  2 eV/$c^2$ neutrinos next to CDM and X-ray gas, all isothermal, and derive the density profiles for clusters such as Coma. 
Nieuwenhuizen (2009) applies to clusters an isothermal fermion model for a single type of dark matter, cold or not, and also 
  isothermal for the galaxies and the X-ray gas. Fitting to weak lensing data of the Abell 1689 cluster works well and yields as the mass 
  $(12/\bar g)^{1/4}\, 1.455 \pm 0.030 \,h_{70} ^{1/2}$ eV/$c^2$, where $\bar g$ is the fermion degeneracy. 
  The only known candidate for such a low mass is the neutrino,  for which $\bar g=3\ast 2\ast2 =12$ modes are available, 
  due to 3 families of (anti)particles with spin $\frac{1}{2}$, because helicity is not conserved in the gravitational field of the cluster.
The mass lies below the experimental upper bound of 2.0 eV/$c^2$ but within the KATRIN window. 
  With this mass active neutrinos constitute about 10\% dark matter. Some 20\% neutrino dark matter may arise provided the 
  sterile partners have been created, too, in the early Universe. This may occur in the presence of a meV-valued Majorana 
  mass matrix, in the time window between the freeze out of quarks and the freeze out of electrons and positrons (Nieuwenhuizen 2009).

  
Umetsu and Broadhurst (2008) combine weak lensing data with reddening of background galaxies out to 3 Mpc, presenting
  the best (extended) weak lensing data to date. Combining them with certain strong lensing data they fit an NFW profile and find a concentration 
  parameter $c_{200} = 13.5$, well exceeding the expected value around 6-7. Nieuwenhuizen and Morandi (2011) employ the 
 Umetsu and Broadhurst (2008) weak lensing data but use better strong lensing data of Limousin et al. (2007) and arrive at the more expected 
 $c_{200} = 6.8$.  However, they point at a 
  systematic deviation in the weak lensing fit, related to the fact that weak lensing alone would give a value of around 7.6 (Limousin et al. 2007).
  They also point at a mild tendency of the strong lensing data to fall well below NFW beyond 300 kpc. 
  
  Coe et al. (2010) present improved strong lensing data by employing a non-parametirc approach.
    Combining these strong lensing data up to 220 kpc with the   weak lensing data of Umetsu and Broadhurst (2008) beyond 220 kpc, 
    they fit to an NFW profile and get $r_s = 338$ kpc, $c_{200} = 7.6$  (or $c_{\rm vir} = 9.6$), 
  leading to r $M_{200} = 1.8^{+0.4}_{-0.3} 10^{15} h_{70}^{-1}M_\odot$ or $M_{\rm vir} = 2.0^{+0.5}_{-0.3} 10^{15} h_{70}^{-1} M_\odot $.
   
    \begin{figure}[h!]
\centering
\includegraphics[width=10cm]{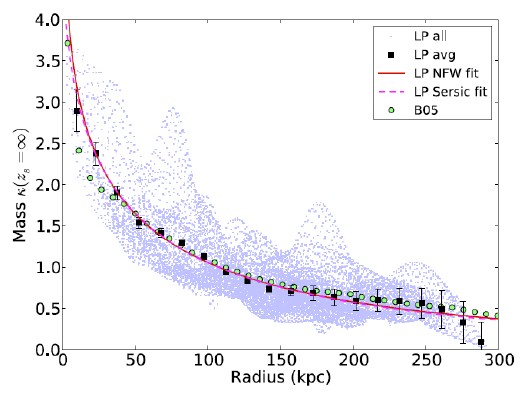}
\caption{Reproduction of Figure 2 of Coe et al. (2010) by permission of the AAS. 
Blue: Strong lensing data of the cluster A1689 obtained with their analysis.
Black: binned data. Green: data by Broahurst et al. (2005).
Red: The NFW model fits well, but it does not support the fast fall off beyond 275 kpc. 
}
  \end{figure}

Nieuwenhuizen and Morandi (2011) also apply the isothermal neutrino model to their A1689 data. The approach again works well, with slightly modified 
  parameters and neutrino mass $1.499 \pm 0.039\,h_{70} ^{-1/2}$ eV/$c^2$. This theory supports a fast decay of strong lensing beyond 300 kpc. 
  The reason is that its neutrino dark matter would be localized up to, say, 300 kpc (Nieuwenhuizen 2009), with 
  the X-ray gas more concentrated in the outskirts.
As seen in Fig 4 for a slightly different data set, in this model strong lensing decays fast beyond the region where DM is localized. Weak lensing, 
  on the other hand, does not exhibit this, as it senses a slow decay, namely an isothermal $1/r$ decay plus an $1/r^2$ correction 
  (Nieuwenhuizen and Morandi, 2011), see also Fig. 5 for weak lensing in the data set considered below.

The case of neutrino dark matter in clusters is also considered by Sanders (2007).
While he assumes modifications of Newton gravity, we shall stay within the Newtonian scope. 
  
\section{Fit to the most recent data sets}
  
We now consider the best data sets to date, the strong lensing of Coe et al. (2010), combined with the weak lensing data of 
Umetsu and Broadhurst (2008) and the X-ray data of Nieuwenhuizen and Morandi (2011).
Whereas Coe et al. (2010) used the strong lensing data up to 220 kpc and the weak lensing data beyond this, we shall now follow our 2011 work and use all 
  strong lensing and weak lensing data, taking the gas data as input. 
  We wish to minimize $\chi^2=\chi^2_{\rm SL} +\chi^2_{\rm WL}$ 
  for the 20 strong lensing data points and 11 weak lensing points.
  
Let us first discuss the NFW case, which involves 2 fit parameters, leaving a minimization over 20+11-2=29 non-trivial inputs. 
(Not having the correlation matrix between the data points of SL or WL, we perform $\chi^2$ fits, here and further on.)
The fit leads to $\chi^2_{\rm SL}= 21.4$ and $\chi^2_{\rm WL}= 14.6$.
The profile parameters are $r_s=291$ kpc and $c_{200} = 8.8$,  or the virial value $c_{\rm vir} = 11.0$. 
With its  $\chi^2/29=1.24$, this fit is satisfactory. 
But, as seen in Fig 4, the strong lensing strongly falls down beyond $r_s$, while the NFW profile continues gradually.
So if the fall off is statistically relevant,  it will pose a problem for the NFW model.

One can also make a fit to the isothermal fermion model assuming an isothermal distribution
for the density of matter in galaxies,
$\rho_G(r)=\bar gm\lambda_T^{-3}\exp[\alpha_G-\beta_Gm\varphi(r)]$  (Nieuwenhuizen 2009).
The fit involves five parameters, the best values being

\BEQ
\label{data5=}
&& \alpha = 29.26\pm 0.43,\quad
\frac{\gb m}{\lambda_T^3}=4092\pm \,275\frac{m_N}{{\rm m}^3},\quad 
\sigma_\nu =564\pm 9\,\frac{{\rm km}}{{\rm s}},\quad  \nn\\&&
\alpha_G=7.44\pm 0.11 ,\quad
\beta_G=\frac{0.745 \pm0.03}{m\sigma_\nu ^2},
\EEQ
where $\sigma_\nu =\sqrt{k_BT/m}$ is a DM velocity dispersion for which $\lambda_T=\hbar\sqrt{2\pi}/m\sigma_\nu$. 
The result for $\beta_G$ differs the most with respect to fits (Nieuwenhuizen and Morandi, 20011), 
but it still is close to $1/m\sigma_\nu^2$, which is the case of virial
equilibrium of galaxies and DM (Nieuwenhuizen 2009). The mass comes out as

\BEQ m=\left(\frac{12}{\bar g}\right)^{1/4}1.55\, \eV \pm 0.04 \,\eV/c^2.\EEQ
From now on, we take $\bar g=12$; in general the masses should be 
 multiplied by the factor $(12/\bar g)^{1/4}$.

The strong lensing fit, shown in Fig. 4, corresponds to $\chi^2_{\rm SL}=57.9$. 
Its match below 100 kpc could be improved by a different modeling of the galaxy density, 
which would also allow to improve the fit between 100 and 300 kpc, so this could easily bring down the present large value
of $\chi^2_{\rm SL}$.   The weak lensing fit is shown in Fig. 5.
The outlying 8'th weak lensing point alone contributes an amount 10.0 to $\chi^2_{\rm WL}=18.1$, the 10 other points bringing 8.1,
so that weak lensing is generally fit quite well.
In total, the present fit thus gives the value $\chi^2/(31-5)=2.92$, larger that for NFW.
We shall not attempt to improve the fit,  since this will have little impact on the value of the neutrino mass.
Moreover, it is seen in Fig. 4 that the neutrino model predicts a fast decay of strong lensing beyond 275 kpc, while NFW does not.
So if this decay of the Coe  et al. (2010) data in Fig. 3 is statistically relevant, it supports the neutrino model and disfavors NFW.
In this regard one may determine till which radius NFW predicts lensing arcs, and compare this with the observations.

\begin{figure}[h!]
\centering
\includegraphics[width=10cm]{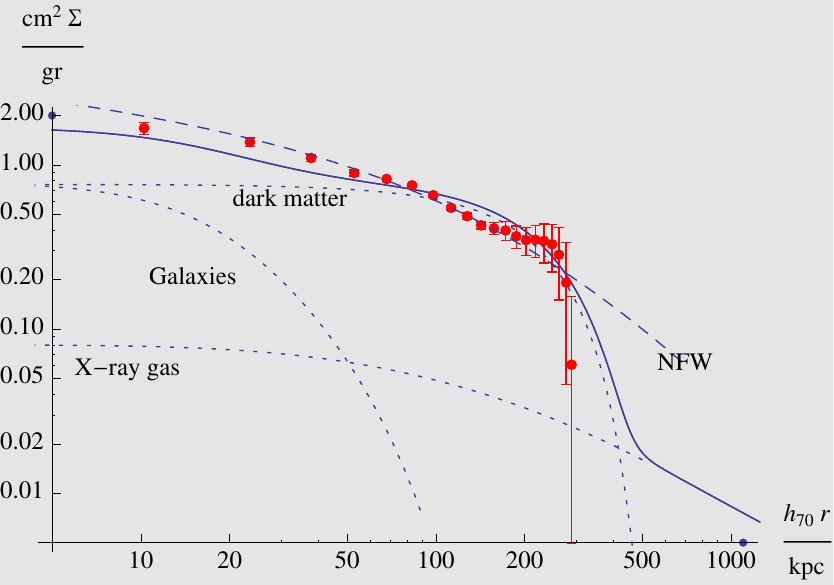}
\caption{Dots: log-log plot of the binned strong lensing data of Coe et al. (2010), see Fig. 3. 
Dashed line: The NFW model fits well, but fails to explain the fast fall off 
  beyond 275 kpc. Full line: Isothermal dark fermions with the measured gas and isothermal modeling of galaxies fit reasonably. 
  Dotted lines: the separate contributions from (neutrino) dark matter, Galaxies and X-ray gas. }
\end{figure}

\begin{figure}[h!]
\centering
\includegraphics[width=10cm]{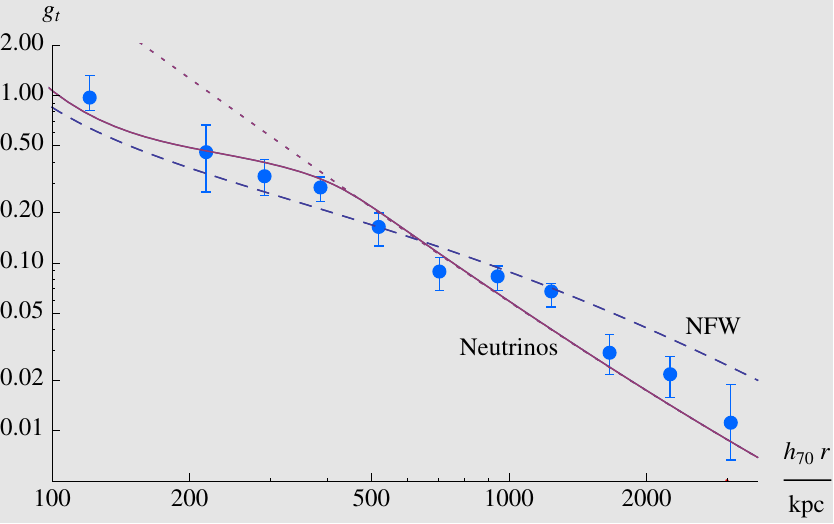}
\caption{Weak lensing data of Umetsu and Broadhurst (2008). Full line: isothermal neutrinos. Dotted line: its asymptotic shape. 
  Dashed line: NFW profile of cold dark matter.}
\end{figure}


\section{Conclusion}

The lensing data of the galaxy cluster Abell 1689 may be explained by isothermal fermions. The best case is neutrinos of nearly 
  the same mass. Combining with previous (but less good) data sets on the same cluster we infer a mass is 
  $m_\nu=1.50 \pm 0.05$ (sys) $\pm 0.03$ (stat) eV/$c^2$. 
 The three species of active neutrinos have been created in thermal equilibrium and 
  they are predicted have total mass of  $M_\nu=3m_\nu\approx 4.5$ eV/$c^2$. With the general relation $\Omega_\nu= M_\nu /(46h_{70}^2)$ 
  the mass density in active neutrinos would be $\Omega^{\rm act}_\nu\approx 10\% $ of the critical mass density of the Universe.
This is much more than the 1.2--3.0\% allowed by the standard $\Lambda$CDM approach under various conditions 
and combinations with other data sets  (Komatsu et al. 2010).  In our philosophy neutrinos should explain most, if not all,
 cluster dark matter, $\Omega_\nu=20-25\%$. 
Under reasonable  conditions,  another 10\% can have arisen from the creation of three 
thermal sterile neutrinos  species of 1.5 eV/$c^2$  (Nieuwenhuizen, 2009).  
If there are more sterile than active partners that would explain even more dark matter.

 The 1.5 eV/$c^2$ mass lies in the range covered by the KATRIN experiment, searching the mass of the electron antineutrino,
  that is expected to start taking data in 2011 (Weinheimer 2009). A discovery around 1.5 eV/$c^2$ would clearly prove our thesis. 
   But non-detection would not necessarily rule out the case of nearly all non-baryonic dark matter in neutrinos.  
  What KATRIN measures is, in leading approximation, 
  the combination $m_{\bar\nu_e}^2\equiv\sum_i|U_{ei}|^2m_i^2$, where $U_{ei}$ is the amplitude of the neutrino mass eigenstate 
  $|\nu_i\rangle$ in the electron antineutrino wavefunction, $|\bar\nu_e\rangle=\sum_iU_{ei}|\nu_i\rangle$,
  with $i$ running from $1$ to $3+N_s$ with $N_s$ the number of sterile neutrinos, and $m_i$ the mass of the eigenstate $|\nu_i\rangle$ 
  (Otten and Weinheimer, 2009). The LSDN  (Aguilar et al. 2001) and MiniBoone data for short baseline neutrino oscillation experiments
  (Aguilar et al. 2010) can be explained by assuming one or more sterile neutrinos with mass of order 1 eV/$c^2$,  
  with the solar and atmospheric  neutrinos involved with much lighter masses.
Therefore a conspiracy may take place, the heavier mass eigenstates having small matrix elements $U_{ei}$, 
making $m_{\bar\nu_e}^2$ too small to be detected at KATRIN,
while dark matter may still come from neutrinos, namely the sterile ones, with masses of order of eV/$c^2$. 
A more interesting scenario is that the masses $m_i$ are well enough separated for KATRIN to observe their effect one-by-one.
  (Otten and Weinheimer, 2009). Such questions and many related ones are beyond the scope of the present paper, 
  but see Nieuwenhuizen (2011).
Clearly, the (sterile) neutrino sector of the standard model of elementary particles is an exciting field of research for the near future.

  As generally expected, the lensing data in A1689 can also be explained by an NFW profile. One may wonder whether there is a discriminating 
  feature between NFW and isothermal neutrinos, that is, between CDM and $\nu$DM. There indeed is a candidate in the fast fall off 
  of strong lensing beyond 270 kpc, first pointed at by Nieuwenhuizen and Morandi (2011) and more convincingly present in the data 
  of Coe et al. (2010), see Figs. 3, 4. 
  NFW would predict a smooth density profile and smooth continuation of strong lensing data. The observations conflict with this and give better support 
for the isothermal neutrino model, where this fall off is related to the fast fall off of neutrino density beyond, say 300 kpc.
The fact that neutrinos are more localized than the X-ray gas also explains the ``missing baryon paradox'' for this cluster, because the
X-ray gas is generally more concentrated in the outskirts, as is concluded from observations on a set of clusters (Rasheed et al 2010). 
  One may investigate whether an NFW profile predicts beyond 300 kpc  more lensing arcs  than the few observed ones.
The answer to this question may touch the validity of the NFW model and its assumed cold dark matter.

  \acknowledgements
  
   A.M. acknowledges support by Israel Science Foundation grant 823/09.

\newcommand{\MNRAS}{Monthly Notices of the Royal Astronomical Society}
  

\subsection*{References}

Evidence for neutrino oscillations from the observation of $\bar\nu_e$  appearance in a $\bar\nu_\mu$  beam,
Physical Review D 64, 112007 [22 pages].

Event Excess in the MiniBooNE Search for $\bar \nu_\mu\rightarrow \bar \nu_e$ Oscillations,
Physical Review Letters 105, 181801 [4 pages].

Akerib, D. S. et al. (CDMS Collaboration) (2010) 
{A low-threshold analysis of CDMS shallow-site data.} 
Physical Review D 82, 122004 [18 pages]

 Altarelli G. and Feruglio F.  (2010).
 Discrete flavor symmetries and models of neutrino mixing,  Review of Modern Physics, 82,  2701Ð2729.
 
Amsler C.,  Doser M., Antonelli M. et al. (Particle Data Group) (2008).  Review of particle physics, Physics Letters  B, 667, 1-1340.

Aprile E., Arisaka K., Arneodo F. et al.  (Xenon 100 Collaboration) (2010). 
First dark matter results from the XENON100 experiment,  Physical Review Letters,  {105}, 131302 [4 pages].

Avignone F. T., Elliott S. R. and Engel J. (2008).
Double beta decay, Majorana neutrinos, and neutrino mass, Review of Modern Physics, 80, 481Ð516.
  
Rasheed B., Bahcall N. and Bode P. (2010).
Where are the missing baryons in clusters?
arXiv:1007.1980 [10 pages].

  Bertone G. (2010), The moment of truth for WIMP Dark Matter, Nature, 468, 389-393.

Cline J. M. (1992).
Constraints on almost-Dirac neutrinos from neutrino-antineutrino oscillations,
Physical Review Letters, 68, 3137-3140.

CMS Collaboration (2011).  
Search for supersymmetry in p-p collisions at 7 TeV in events with jets and missing transverse energy,
arXiv:1101.1628 [30 pages].
  
Coe D., Narciso B., Broadhurst T. and Moustakas L. A. (2010)
A high-resolution mass map of galaxy cluster substructure: LensPerfect analysis of A1689,
Astrophysical Journal, 723, 1678-1702.

Cowsik, R. and McClelland, J. (1973). Gravity of neutrinos of nonzero mass in astrophysics, 
Astrophysical Journal, 180, 7-10.

Feng J.L. (2010). Dark Matter Candidates from Particle Physics and Methods of Detection, 
Annual Review of Astronomy and Astrophysics, 48, 495-545.

Gibson C. H. and Schild R. E.  (2011). Primordial planet formation, Journal of Cosmology .

Gibson G. H. (2011).  Why the dark matter of galaxies is clumps of micro-brown dwarfs
and not Cold Dark Matter, Journal of Cosmology .

Komatsu E., Smith K. M., Dunkley J. et al. (2010).
Seven-year Wilkinson anisotropy probe (WMAP) observations: Cosmological interpretation,
Astrophysical Journal Supplement Series, 192, 18 [47pp].

 Kraus Ch., Bornschein B. et al. (2005). Final Results from phase II of the Mainz Neutrino Mass Search in Tritium beta-Decay,
  EuroPhysical Journal C, 40,  447-468. 

 Kusenko A. (2009). Sterile neutrinos: The dark side of the light fermions,  Physics Reports, 481, 1-28.

Lesgourgues J. and Pastor S. (2006). Massive neutrinos and cosmology, Physics Reports, 429, 307-379.

Limousin  M. et al. (2007). Combining Strong and Weak Gravitational Lensing in Abell 1689,
 Astrophysical Journal  { 668}, 643-666.

Lobashev V. M., Aseev V. M.  et al. (1999). Direct search for mass of neutrino and anomaly in the tritium beta-spectrum,
Physics Letters  B, 460, 227-235.

 Lynden Bell D. (1967). Statistical mechanics of violent relaxation in stellar systems,
 \MNRAS \ 136, 101-121.
  
Navarro, J. F., Frenk, C. S. and White, S. D. M. (1997).  A universal density profile from hierarchical clustering,
Astrophysical Journal 490, 493-508.
  
Nieuwenhuizen, T. M.  (2009). 
\emph{}{Do non-relativistic neutrinos constitute the dark matter?} 
Europhysics Letters, 86, 59001 [6 pages].

Nieuwenhuizen, T. M.  (2011). 
Sterile neutrinos: their genesis, oscillations  and role in cosmology,
Journal of Cosmology .

Nieuwenhuizen, T. M., Gibson, C. H. and Schild, R. E. (2009).
{Gravitational hydrodynamics of large-scale structure formation},
{ Europhysics Letters},  {88}, 49001 [6 pages].

Nieuwenhuizen, T. M.  and Morandi M. (2011). 
Are observations of the galaxy cluster Abell 1689 consistent with a neutrino dark matter scenario?
submitted [6 pages].

Otten, E. W. and  Weinheimer C. (2008).
Neutrino mass limit from tritium beta decay, Reports on Progress in Physics, 71, 086201 [81 pages].

  Rasheed, B., Bahcall, N. and Bode, P. (2010).  Where are the missing baryons in clusters?  arXiv:1007.1980 [10 pages].


Russell, D. G., (2005a) Further Evidence for Intrinsic Redshifts in Normal Spiral Galaxies. Astrophysics and Space Science, 299, 2005, 387-403.

Sanders R. H. (2007).  Neutrinos as cluster dark matter, \MNRAS, 380, 331-338.

Umetsu, K. and Broadhurst, T. (2008).  
Combining lens distortion and depletion to map the mass distribution of A1689, 
Astrophysical Journal,  684, 177-203.

  Weinheimer, C. (2009). Direct determination of Neutrino Mass from Tritium Beta Spectrum. 
{arXiv}:0912.1619 [13 pages].


\end{document}